\documentclass{article}

\usepackage{amssymb,amsmath}
\usepackage{epsfig}

\usepackage{color}
\usepackage{amsmath}
\usepackage{amssymb}
\usepackage{amsthm}

\newcommand\x{{\bf x}}
\newcommand\y{{\bf y}}
\newcommand\z{{\bf z}}
\newcommand\bfa{{\bf a}}
\newcommand\bfb{{\bf b}}

\newcommand\bfq{{\bf q}}

\newcommand\zero{{\bf 0}}
\newcommand\cc{{\mathbb C}}

\begin{document}

\title{GPU acceleration of Newton's method \\
       for large systems of polynomial equations \\
       in double double and quad double arithmetic}
%       in double double and quad double arithmetic\thanks{This
%material is based upon work supported by the National Science Foundation
%under Grant No.\ 1115777.} }

\author{Jan Verschelde and Xiangcheng Yu \\
Department of Mathematics, Statistics, and Computer Science \\
University of Illinois at Chicago, 851 South Morgan (M/C 249) \\
Chicago, IL 60607-7045, USA \\
emails: {\tt jan@math.uic.edu} and {\tt xyu30@uic.edu}}

\date{12 May 2014}

\maketitle

\begin{abstract}
In order to compensate for the higher cost of double double 
and quad double arithmetic when solving large polynomial systems, 
we investigate the application of NVIDIA Tesla K20C
general purpose graphics processing unit.
The focus on this paper is on Newton's method, which requires the
evaluation of the polynomials, their derivatives, and the solution
of a linear system to compute the update to the current approximation
for the solution.
The reverse mode of algorithmic differentiation for a product of
variables is rewritten in a binary tree fashion so all threads in
a block can collaborate in the computation.
For double arithmetic, the evaluation and differentiation problem
is memory bound, whereas for complex quad double arithmetic the
problem is compute bound.
With acceleration we can double the dimension and get results that
are twice as accurate in about the same time.

\medskip

\end{abstract}

\section{Introduction}

We investigate the application of general purpose graphics processing
units (GPUs) to solving large systems of polynomial equations with
numerical methods.  Large systems not only lead to an increased
number of operations, but also to more accumulation of numerical roundoff
errors and therefore to the need to calculate in a precision that is higher
than the common double precision.  Motivated by the need of higher numerical
precision, we can formulate our goal more precisely.  With massively
parallel algorithms we aim to offset the extra cost of double double
and quad double arithmetic~\cite{HLB00,LHL10} 
and achieve quality up~\cite{Akl04}, a project we started in~\cite{VY10}.

\noindent {\bf Problem Statement.}  Our problem is to
accelerate Newton's method for large polynomial systems, aiming to offset
the overhead cost of double double and quad double complex arithmetic.
We assume the input polynomials are given in their sparse
distributed form: all polynomials are fully expanded and only those
monomials that have a nonzero coefficient are stored.
For accuracy and application to overdetermined systems, 
we solve linear systems in the least squares sense
and implement the method of Gauss-Newton.

Our original massively parallel algorithms for evaluation and differentiation
of polynomials~\cite{VY12} and for the modified Gram-Schmidt
method~\cite{VY13} were written with a fine granularity, making intensive 
use of the shared memory.  The limitations on the capacity of the shared 
memory led to restrictions on the dimensions on the problems we could solve.
These problems worsened for higher levels of precision, in contrast
to the rising need for more precision in higher dimensions.

\noindent {\bf Related Work.}  As the QR decomposition is of fundamental
importance in applied linear algebra many parallel implementations have
been investigated by many authors, see e.g.~\cite{AADFLTT11}, \cite{ABDK11}.
A high performance implementation of the QR algorithm 
on GPUs is described in~\cite{KCR09}.  
In~\cite{BASR12}, the performance of CPU and GPU implementations
of the Gram-Schmidt were compared.
A multicore QR factorization is compared to a GPU implementation
in~\cite{LD11}.
GPU algorithms for approaches related to QR and Gram-Schmidt
are for lattice basis reduction~\cite{BG12} and 
singular value decomposition~\cite{FMW12}.
In~\cite{VD08}, the left-looking scheme is dismissed because
of its limited inherent parallelism and as in~\cite{VD08}
we also prefer the right-looking algorithm for more thread-level
parallelism.

The application of extended precision to BLAS 
is described in~\cite{BLAS02},
see~\cite{DHLR09} for least squares solutions. 
The implementation of BLAS routines on GPUs in
triple precision (double + single float) is discussed in~\cite{MT12}.
In~\cite{Rum10}, double double arithmetic is described under the
section of error-free transformations.
An implementation of interval arithmetic on CUDA GPUs is
presented in~\cite{CDD12}.

The other computationally intensive stage in the application
of Newton's method is the evaluation and differentiation of the system.
Parallel automatic differentiation techniques are
described in~\cite{BGKW08}, \cite{GPGK08}, and~\cite{UHHHHN09}.

Concerning the GPU acceleration of polynomial systems solving,
we mention two recent works.
A subresultant method with a CUDA implementation of the FFT 
to solve systems of two variables is presented in~\cite{MP11}.
In~\cite{KP12}, a CUDA implementation for an NVIDIA GPU
of a multidimensional bisection algorithm is discussed.

\noindent {\bf Our contributions.}  
For the polynomial evaluation and differentiation we reformulate
algorithms of algorithmic differentiation~\cite{GW08} applying 
optimized parallel reduction~\cite{Harris} to the products that 
appear in the reverse mode of differentiation.
Because our computations are geared towards
extended precision arithmetic which carry a higher cost per operation,
we can afford a fine granularity in our parallel algorithms.
Compared to our previous GPU implementations in~\cite{VY12,VY13},
we have removed the restrictions on the dimensions and are now able
to solve problems involving several thousands of variables.
The performance investigation involves mixing the memory-bound
polynomial evaluation and differentiation with the compute-bound
linear system solving.

\section{Polynomial Evaluation and Differentiation}

We distinguish three tasks in the evaluation and differentiation
of polynomials in several variables given in their sparse
distributed form.  First, we separate the high degree parts
into common factors and then apply algorithmic differentiation
to products of variables.  In the third stage, monomials are
multiplied with coefficients and the terms are added up.

\subsection{common factors and tables of power products}
A monomial in $n$ variables is defined by a sequence of 
natural numbers $d_i \geq 0$, for $i = 1,2,\ldots,n$.
We decompose a monomial as follows:
\begin{equation}
   x_1^{d_1} x_2^{d_2} \cdots x_n^{d_n}
   = x_{i_1} x_{i_2} \cdots x_{i_k}
   \times
   x_{j_1}^{e_{j_1}} x_{j_2}^{e_{j_2}} \cdots x_{j_\ell}^{e_{j_\ell}}
\end{equation}
where $x_{i_1} x_{i_2} \cdots x_{i_k}$ is the product of all $k$
variables that have a nonzero exponent.
The $\ell$ variables that appear with a positive exponent
occur in
$x_{j_1}^{e_{j_1}} x_{j_2}^{e_{j_2}} \cdots x_{j_\ell}^{e_{j_\ell}}$
with exponent $e_{j_i} = d_{j_i} - 1$, for $i = 1,2,\ldots, \ell$.

We call the monomial
$x_{j_1}^{e_{j_1}} x_{j_2}^{e_{j_2}} \cdots x_{j_\ell}^{e_{j_\ell}}$
a {\em common factor}, as this factor is a factor in all partial
derivatives of the monomial.
Using tables of pure powers of the variables, the values of the 
common factors are products of the proper entries in those tables.
The cost of evaluating monomials of high degrees is thus deferred
to computing powers of the variables.
The table of pure powers is computed in shared memory 
by each block of threads.

\subsection{evaluation and differentiation of a product of variables}
Consider a product of variables: $x_1 \star x_2 \star \cdots \star x_n$.
The straightforward evaluation and the computation of the gradient
takes $n-1 + n \times (n-2) = n^2 - n - 1$ multiplications.
Recognizing the product as the example of Speelpenning in algorithmic
differentiation~\cite{GW08}, 
the number of multiplications to evaluate the product
and compute all its derivatives drops to~$3n-5$.

The computation of the gradient requires in total $n-1$ 
extra memory locations.  We need $n-2$ locations for the intermediate 
forward products $x_1 \star x_2 \star \cdots \star x_k$, 
$k=2, 3, \ldots, n-1$.
For the backward products $x_n \star x_{n-1} \star \cdots \star x_k$, 
$k = n-1, n-2, \ldots, 3$ only one extra temporary memory location
is needed, as this location can be reused each time for the next
backward product, if the computation of the backward products is
interlaced with the multiplication of the forward with the 
corresponding backward product.

\begin{figure}[h!]
\begin{center}
\begin{picture}(350,100)(0,0)
\put(0,0){
\begin{picture}(80,80)(0,0)
\put(0,0){$x_1$}
\put(25,0){$x_2$}
\put(50,0){$x_3$}
\put(75,0){$x_4$}
\put(2,30){$x_1 \star x_2$}
\put(5,8){\vector(1,2){9}}
\put(28,8){\vector(-1,2){9}}
\put(52,30){$x_3 \star x_4$}
\put(55,8){\vector(1,2){9}}
\put(78,8){\vector(-1,2){9}}
\put(17,70){$x_1 x_2 \star x_3 x_4$}
\put(25,40){\vector(1,2){12}}
\put(58,40){\vector(-1,2){12}}
\end{picture}
}
\put(140,0){
\begin{picture}(200,80)(0,0)
\put(0,70){$x_1 \star x_3 x_4$}
\put(50,70){$x_2 \star x_3 x_4$}
\put(100,70){$x_1 x_2 \star x_3$}
\put(150,70){$x_1 x_2 \star x_4$}
\put(0,0){$x_1$}
\put(5,10){\vector(0,1){55}}
\put(50,0){$x_2$}
\put(55,10){\vector(0,1){55}}
\put(130,0){$x_3$}
\put(135,10){\vector(0,1){55}}
\put(180,0){$x_4$}
\put(185,10){\vector(0,1){55}}
\put(150,30){$x_3 x_4$}
\put(158,38){\vector(-3,1){78}}
\put(150,36){\vector(-4,1){112}}
\put(20,30){$x_1 x_2$}
\put(28,38){\vector(3,1){78}}
\put(38,36){\vector(4,1){112}}
\end{picture}
}
\end{picture}
\caption{The left picture shows
the evaluation of a product of variables organized in a binary tree,
starting at the leaves and placing the result at the root of the tree.
The picture at the right 
shows the computation of all derivatives (the gradient), 
with inputs taken from different levels in the tree.
We count 3 $\star$ to evaluate and 4 more $\star$ to differentiate.}
\label{figcircuit1}
\end{center}
\end{figure}
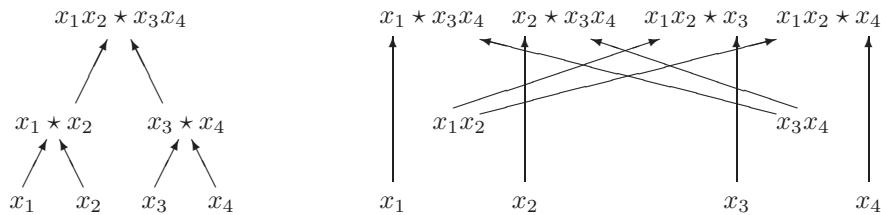

For $n = 4$,
Figure~\ref{figcircuit1} displays two arithmetic circuits,
one to evaluate a product of variables and another to compute its gradient.
The second circuit is executed after the first one, using the same
tree structure that holds intermediate products.
At a node in a circuit, we write $x_1 \star x_2$ 
if the multiplication $\star$ happens at the node
and we write $x_1 x_2$ if we use the value of the product.
At most one multiplication is performed at each node of the circuit.

\begin{figure*}[t!]
\begin{center}
\begin{picture}(455,160)(0,0)
\put(0,0){$x_1$}     \put(5,10){\vector(0,1){134}}
\put(60,0){$x_2$}    \put(65,10){\vector(0,1){134}}
\put(120,0){$x_3$}   \put(125,10){\vector(0,1){134}}
\put(180,0){$x_4$}   \put(185,10){\vector(0,1){134}}
\put(240,0){$x_5$}   \put(245,10){\vector(0,1){134}}
\put(300,0){$x_6$}   \put(305,10){\vector(0,1){134}}
\put(360,0){$x_7$}   \put(365,10){\vector(0,1){134}}
\put(420,0){$x_8$}   \put(425,10){\vector(0,1){134}}
\put(0,150){$x_1 \star x_{3:8}$}
\put(60,150){$x_2 \star x_{3:8}$}
\put(120,150){$x_3 \star x_{1:2} x_{5:8}$}
\put(180,150){$x_4 \star x_{1:2} x_{5:8}$}
\put(240,150){$x_5 \star x_{1:4} x_{7:8}$}
\put(300,150){$x_6 \star x_{1:4} x_{7:8}$}
\put(360,150){$x_7 \star x_{1:6}$}
\put(420,150){$x_8 \star x_{1:6}$}
\put(85,28){$x_{1:4}$}
\put(105,30){\vector(4,1){267}}
\put(90,38){\vector(3,1){170}}
\put(250,100){$x_{1:4} \star x_{5:6}$}
\put(283,110){\vector(3,1){100}}
\put(295,105){\vector(4,1){145}}
\put(370,100){$x_{1:4} \star x_{7:8}$}
\put(385,110){\vector(-3,1){100}}
\put(395,110){\vector(-2,1){60}}
\put(15,15){$x_{1:2}$}   \put(20,25){\vector(0,1){70}}
\put(135,15){$x_{3:4}$}  \put(140,25){\vector(0,1){70}}
\put(275,15){$x_{5:6}$}
\put(280,25){\vector(0,1){70}}
\put(395,15){$x_{7:8}$}
\put(400,25){\vector(0,1){70}}
\put(325,28){$x_{5:8}$}
\put(320,30){\vector(-4,1){267}}
\put(335,38){\vector(-3,1){170}}
\put(15,100){$x_{1:2} \star x_{5:8}$}
\put(48,110){\vector(3,1){100}}
\put(60,105){\vector(4,1){145}}
\put(135,100){$x_{3:4} \star x_{5:8}$}
\put(148,110){\vector(-3,1){105}}
\put(160,110){\vector(-2,1){65}}
\end{picture}
\caption{In the arithmetic circuit to differentiate a product of 8 
variables, the inputs are the 8 variables: $x_1$, $x_2$, $\ldots$, $x_8$;
the 4 products of 2 consecutive variables:
$x_{1:2}$, $x_{3:4}$, $x_{5:6}$, and~$x_{7:8}$; and the 2 products 
of 4 consecutive variables: $x_{1:4}$ and~$x_{5:8}$. 
The nodes $x_{1:4}$ and~$x_{5:8}$ have $x_{1:8}$ 
(omitted because not used) as their common ancestor, 
which is the root of the tree representation of the
differentiation circuit.  }
\label{figcircuit2}
\end{center}
\end{figure*}
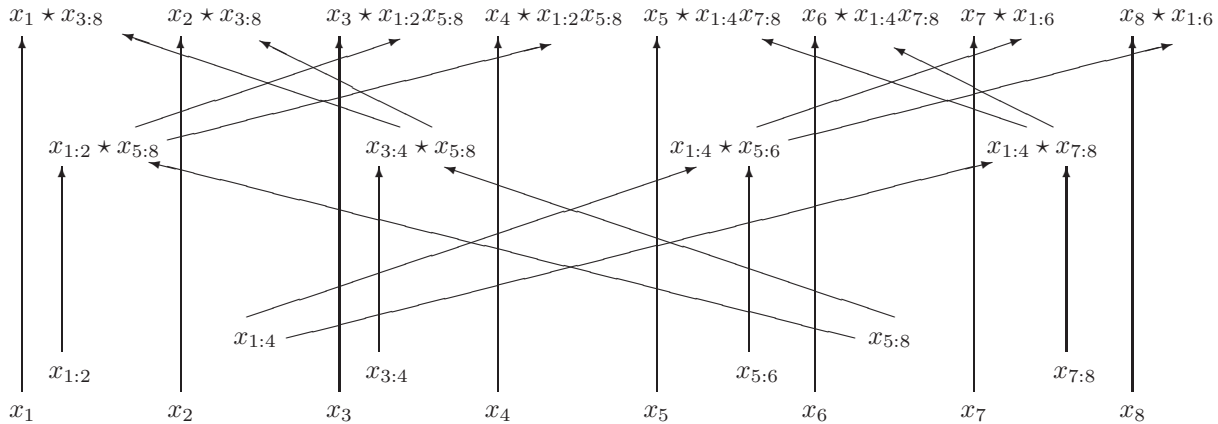

Denote by $x_{i:j}$ the product 
$x_i \star \cdots \star x_k \star \cdots \star x_j$,
for all natural numbers~$k$ between $i$ and $j$.
Figure~\ref{figcircuit2} displays the arithmetic circuit to compute
all derivatives of a product of 8 variables, after the evaluation
of the product in a binary tree.

To count the number of multiplications to evaluate, we restrict to the
case of a complete binary tree, i.e.: $n = 2^k$ for some~$k>0$ and
compute the sum $\displaystyle 
1 + 2 + 4 + \cdots + n/2 = \sum_{i=0}^{n/2} 2^i = n - 1$.
The circuit to compute all derivatives contains a tree of the same size:
with $n-1$ nodes, so the number of multiplications equals $n-1$
minus 3 for the nodes closest to the root which require no computations,
and plus $n$ for the multiplications at the leaves: $2n - 4$ in total.
So the total number of multiplications to evaluate a product of $n$ 
variables and compute its gradient with a binary tree equals~$3n - 5$.

While keeping the same operational cost of~$2n - 5$ as the original
algorithm, the organization of the multiplication in a binary tree
incurs less roundoff.  In particular the roundoff error for the
evaluated product will be proportional to $\log_2(n)$ instead of~$n$ 
of the straightforward multiplication.  
For a large number of variables, such as $n = 1,024$,
this reorganization improves the accuracy by two decimal places.
The improved accuracy of the evaluated product does not cost more 
storage as the size of binary tree equals $n-1$.

For the derivatives, the roundoff error is bounded by the number
of levels in the arithmetic circuit, which is
\newline $\displaystyle \sum_{k=1}^{\log_2(n)} k 
= \frac{1}{2} (\log_2(n))^2 + \log_2(n)$.
While this bound is still better than~$n-1$,
the improved accuracy for the gradient comes at the extra cost
of~$n-4$ additional memory locations, needed as nodes in the arithmetic
circuit for the gradient.
In shared memory, the memory locations for the input variables $x_k$
are overwritten by the corresponding components of the gradient,
e.g.: $x_1 \star x_2 \cdots x_n$ 
then occupies the location of~$x_1$. 

In the original formulation of the computation of the example
of Speelpenning, only one thread performed all computation for
one product and the parallelism consisted in having enough
monomials in the system to occupy all threads working separately
on different monomials.  The reformulation of the evaluation
and differentiation with a binary tree allows for several threads
to collaborate on the computation of one large product.
The reformulation refined the granularity of the parallel algorithm
and we applied the techniques suggested in~\cite{Harris}.

If $n$ is not a power of 2, then for some positive $k$ and~$\ell$,
denote $n = 2^k + \ell < 2^{k+1}$.
The first $\ell$ threads load two variables and are in charge of
the product of those two variables, while other threads load
just one variable.  
The multiplication of values for variables of consecutive index,
e.g.: $x_1 \star x_2$ will result in a bank conflict in shared memory
as threads require data from an even and odd bank.
To avoid bank conflicts, the computations are rearranged,
e.g. as $(x_1 \star x_3) \star (x_2 \star x_4)$,
so thread~0 operates on~$x_1, x_3$ and thread~1 on~$x_2, x_4$.

Table~\ref{tabmoneval} shows the results on the evaluation
and differentiation of products of variables in double arithmetic,
applying the techniques of~\cite{Harris}.
The first GPU algorithm is the reverse mode algorithm that takes
$3n - 5$ operations executed by one thread per monomial.
When all threads in a block collaborate on one monomial
in the second GPU algorithm we observe a significant speedup.
Speedups and memory bandwidth improve when resolving the
bank conflicts in the third improvement.
The best results are obtained adding unrolling techniques.

\begin{table}[bh]
\caption{Evaluation and differentiation of 65,024 monomials
in 1,024 doubles.  
Times on the K20C obtained with {\tt nvprof} (the NVIDIA profiler)
are in milliseconds (ms).
Dividing the number of bytes read and written by the time
gives the bandwidth.
Times on the CPU are on one 2.6GHz Intel Xeon E5-2670,
with code optimized with the {\tt -O2} flag.}
\begin{center}
\begin{tabular}{r|c|rr|r}
    &           method          & time~~~  & bandwidth  & speedup \\ \hline
CPU &                           & 330.24ms &            &         \\ \hline
GPU & one thread per monomial   &  86.43ms &            &   3.82~  \\
    & one  block per monomial   &  15.54ms &  79.81GB/s &  21.25~  \\
    & sequential addressing     &  14.08ms &  88.08GB/s &  23.45~  \\
    &  unroll last wrap         &  10.19ms & 121.71GB/s &  32.40~  \\
    &  complete unroll          &   9.10ms & 136.28GB/s &  36.29~
\end{tabular}
\label{tabmoneval}
\end{center}
\end{table}

In Table~\ref{tabmoneval}, one block of threads computes the value
and the gradient of one product in 1,024 variables.
Instead of one large product, with our code one block can compute 
many monomials of smaller sizes.  In the arithmetic circuits of
Figure~\ref{figcircuit1} and~\ref{figcircuit2}, instead of going
all the way to the root of the tree,
the computation stops at some intermediate level.
Table~\ref{tabmoneval2} present timings for this computation.

\begin{table}[bh]
\begin{center}
\caption{Evaluation and differentiation of $m$ monomials 
of different size~$n$
by 65,024 blocks with 512 threads per block for 1,024 doubles in shared
memory, accelerated by the K20C with timings in milliseconds obtained by
the NVIDIA profiler. 
Times on the CPU are on one 2.6GHz Intel Xeon E5-2670,
with code optimized with the {\tt -O2} flag.}
\label{tabmoneval2}
\begin{tabular}{r| r| r r| r }
\multicolumn{1}{c|}{$n$} & \multicolumn{1}{c|}{$m$}
&  \multicolumn{1}{c}{CPU} & \multicolumn{1}{c|}{GPU}  & speedup \\ \hline
1024   &   1  & 330.24ms  & 9.12ms & 36.20~ \\
 512   &   2  & 328.92ms  & 8.73ms & 37.66~ \\
 256   &   4  & 320.78ms  & 8.84ms & 36.29~ \\ 
 128   &   8  & 309.02ms  & 8.15ms & 37.89~ \\
  64   &  16  & 289.30ms  & 7.27ms & 39.77~ \\
  32   &  32  & 256.07ms  & 9.51ms & 26.94~ \\
  16   &  64  & 230.34ms  & 8.86ms & 25.99~ \\ 
   8   & 128  & 218.74ms  & 7.79ms & 28.07~ \\ 
   4   & 256  & 202.20ms  & 7.05ms & 28.69~ \\
\end{tabular}
\end{center}
\end{table}

The evaluation and differentiation of products of variables
is memory bound for complex double arithmetic and the techniques illustrated
by Table~\ref{tabmoneval} are also relevant 
for real double double arithmetic.
In complex double double and quad double arithmetic, 
the cost overhead of the arithmetic
dominates, the computation becomes compute bound and we use global memory.

\subsection{coefficient multiplication and term summation}

The third task is to multiply every derivative of the product
of variables with the common factor and the proper coefficient,
multiplied with the values of the exponents.
Then a sum reduction of the evaluated terms gives the
values of the polynomials and the Jacobian matrix.
The efficient implementation of the scan with CUDA
is described in~\cite{Harris2}.

\section{Orthogonalization and Delayed Normalization}

Before describing our massively parallel algorithms for the
modified Gram-Schmidt method, % ~\cite{LBG13}, 
we formalize the notations.
We typically compute with complex numbers and follow notations
in~\cite{GV96} for the complex conjugated inner product~$\x^H \y$.
Pseudo code of the modified Gram-Schmidt orthogonalization method
is listed in Figure~\ref{figalgmgs}.
\begin{figure}[hbt]
\begin{center}
\begin{tabbing}
\hspace{8mm} \= Input: $A \in \cc^{m \times n}$. \\
\> Out\=put\=: \= $Q \in \cc^{m \times n}$, $R \in \cc^{n \times n}$: 
                  $Q^H Q = I$, \\
\>    \>   \>  \> $R$ is upper triangular, and $A = QR$. \\
\> let ${\bfa}_k$ be column~$k$ of~$A$ \\
\> for $k$ from 1 to $n$ do \\
\>    \> $r_{k, k} := \sqrt{ \bfa_k^H \bfa_k }$ \\
\>    \> $\bfq_k := \bfa_k/r_{k, k}$, $\bfq_k$ is column $k$ of~$Q$ \\
\>    \> for $j$ from $k+1$ to $n$ do \\
\>    \>   \> $r_{k, j} := \bfq_k^H \bfa_j$ \\
\>    \>   \> $\bfa_j := \bfa_j - r_{k, j} \bfq_k$
\end{tabbing}
\caption{The modified Gram-Schmidt orthogonalization algorithm.}
\label{figalgmgs}
\end{center}
\end{figure}

The modified Gram-Schmidt method computes the 
the QR decomposition of a matrix~$A$,
which allows to solve the linear system~$A \x = \bfb$
in the least squares sense, minimizing~$|| \bfb - A \x ||_2^2$.
In the reduction of $A \x = \bfb$ to an upper triangular system
$R \x = Q^H \bfb$, we do not compute $Q^H \bfb$ separately.
As recommended in~\cite[\S 19.3]{Hig96} for numerical stability 
the modified Gram-Schmidt method is applied to the matrix~$A$
augmented with~$\bfb$:

\begin{equation}
   \left[ 
      \begin{array}{cc}
         A & \bfb
      \end{array}
   \right]
   = 
   \left[
      \begin{array}{cc}
         Q & \bfq_{n+1}
      \end{array}
   \right]
   \left[
      \begin{array}{cc}
         R & \y \\
         0 & z
      \end{array}
   \right].
\end{equation}
Because $\bfq_{n+1}$ is orthogonal to the column space of~$Q$:
$|| \bfb - A \x ||_2^2 = ||R \x - \y||_2^2 + z^2$ and $\y = Q^H \bfb$.
As a check on the correctness and the accuracy of our computed results, 
we wrote code to compute~$A - QR$.
Although computing the entire~$A - QR$, the test whether 
$a_{m,n} - \bfq_n^H {\bf r}_n$ is small enough 
(where ${\bf r}_n$ is the last column of~$R$)
could already indicate whether the working precision was sufficient.

The algorithm in Figure~\ref{figalgmgs} starts with 
the computation of the complex conjugated inner product
$\bfa_k^H \bfa_k$, followed by the normalization
$\bfq_k := \bfa_k/r_{k, k}$, where $r_{k, k} := \sqrt{ \bfa_k^H \bfa_k }$.
For the inner product, we load the components of an $m$-dimensional vector
into shared memory.  Denoting the number of components that fit into the
shared memory by~$K$ (its capacity), then let $L = \lceil m/K \rceil$
be the number of rounds it takes to compute
\begin{equation} \label{eqrounds4ip}
   \bfa_k^H \bfa_k = \sum_{i=0}^{L-1} \sum_{j=0}^{K-1}
   \overline{a}_{k, i K + j} \star a_{k, i K + j},
\end{equation}
where the indexing of the components of a vector starts at zero
and $\overline{a}$ denotes the complex conjugate of $a \in \cc$.
The value for $K$ is typically a multiple of the warp size
and equals the number of threads in a block.
In~(\ref{eqrounds4ip}), the index~$j$ is the index of the thread
in a block, so the inner loop is performed simultaneously in one step
by all threads in the block.  The outside loop on $i$ is done in
a sum reduction and takes $\log_2(L)$ steps.
The computation of $\bfa_k^H \bfa_k$ for an $n$-dimensional vector~$\bfa_k$
is reduced to $m$ memory accesses, $L$ steps to make all partial sums
$\displaystyle \sum_{j=0}^{K-1} 
 \overline{a}_{k, i K + j} \star a_{k, i K + j}$,
and then $\log_2(L)$ steps for the outer sum.

When $\star$ is performed in standard precision with hardware arithmetic,
then~(\ref{eqrounds4ip}) seems to be memory bound,
but for the $\star$ in double double and quad double precision,
performed by arithmetic encoded in the software, the inner product
becomes compute bound as the compute to memory access ratio becomes
large enough to offset memory accesses.

For the reduction, we compute the inner product
$r_{k, j} := \bfq_k^H \bfa_j$ of two $m$-vectors:

\begin{equation}
  \begin{array}{c}
    \bfq_k \\
    \left[
      \begin{array}{c} q_{k,0} \\ q_{k,1} \\ \vdots \\ q_{k,m-1} \end{array}
    \right]
  \end{array}
  ~
  \begin{array}{c}
    \bfa_j \\
    \left[
      \begin{array}{c} a_{j,0} \\ a_{j,1} \\ \vdots \\ a_{j,m-1} \end{array}
    \right]
  \end{array}
  ~
  \begin{array}{c}
    \bfq_k^H \bfa_j \\
    \left[
      \begin{array}{c}
         \bar{q}_{k,0} \star a_{j,0} \\ 
         \bar{q}_{k,1} \star a_{j,1} \\ \vdots \\
         \bar{q}_{k,K-1} \star a_{j,m-1}
      \end{array}
    \right]
  \end{array}
\end{equation}

As we can keep $K$ components of each vector in shared memory,
thread $t$ in a block computes $\bar{q}_{k,t} \star a_{j,t}$.
If we may override $\bfq_k$, then $2 m$ shared memory locations suffice,
but we still need $\bfq_k$ for $\bfa_j := \bfa_j - r_{k, j} \bfq_k$.
In total we need $3 m$ shared memory locations to perform the reductions.
Similar to the inner product for the norm of~$\bfa_k$,
the computation of~$\bfq_k^H \bfa_j$ is performed in $L$ rounds,
where $L = \lceil 3m/K \rceil$, for the capacity $K$ of shared memory.

The calculation of the inner products in $L$ rounds is the first
modification to our original massively parallel Gram-Schmidt implementation.
The second modification is the delay of the normalization.
In the next paragraph we explain the need for this delay.

In the reduction stage, the inner $j$-loop is executed by $n-k$ blocks
of threads.  Every block of threads performs the normalization of the
$k$-th pivot column before proceeding to the reduction.
The first block writes the normalized vector into global memory,
all other blocks write the reduced vectors into global memory.
In the new revised implementation, each vector is processed in
several rounds and is read from global memory into shared memory
not only at the beginning of the calculations.
For large dimensions, not all blocks can be launched simultaneously.
It may even be that the block that will reduce the last column
is not even scheduled for launching at a time when the first block
has finished its writing of the normalized $\bfa_k$ into global memory.

As some block would load in (partially) normalized vectors in the
reduction stage, we propose to delay the normalization to the next
iteration of the $k$-loop in the algorithm in Figure~\ref{figalgmgs}.  
At each iteration, the first block writes the norm of the current 
pivot column to a location in global memory and normalizes the previous
pivot column, dividing every component of the previous pivot column 
by its norm stored in global memory and writes then the normalized 
previous column into global memory. 
With delayed normalization, the column $\bfq_k$ is computed last
and is only stored in step~$k+1$.
At the very end of the algorithm, there is one extra kernel launch
for the normalization that leads to~$\bfq_n$.

The application of shared memory to reduce global memory traffic
is referred to as tiling~\cite[pages 108-109]{KH13}.
Our tiles consist of slices of one column as we assign one column
to one block.  If we want to reduce the number of kernel launches,
we could assign multiple (adjacent) columns to one block
to make proper tiles as submatrices of the original matrix.

Our implementation of the modified Gram-Schmidt method is a
right-looking algorithm, as this gives the most thread-level
parallelism, pointed out in~\cite{VD08}.
Using a right-looking algorithm, we launch as many blocks as
there are columns to update, where each block can work on one column.
The cost of memory traffic is mitigated with shared memory
and for double double and quad double precision, the cost of
the software arithmetic dominates the cost of memory accesses
so good speedups are obtained over optimized serial code.

The third modification concerns the back substitution 
to compute the least squares solution to~$R x = Q^H \bfb$.
Limited by the capacities of shared memory in our previous implementation
only one block of threads performed the back substitution.
For larger dimensions, denoting $Q^H \bfb$ by $\y$, the computation of
\begin{equation} \label{eqbackrounds}
   r_{\ell, \ell} x_\ell = y_\ell - \sum_{j=0}^{\ell - 1} r_{\ell, j} x_j
   = y_\ell - \sum_{i=0}^{L-1} \sum_{j = 0}^{K-1} r_{\ell, i K + j},
\end{equation}
where $L = \lceil m/K \rceil$, for the capacity $K$ of shared memory.
The main difference with our previous implementation is that now $L$
blocks can work simultaneously at the evaluation of various components
of~(\ref{eqbackrounds}).  The pivot block computes the actual components
of the solution, while the other blocks compute the reductions for
components at the low indices and write the reductions of the right
hand side vector into global memory for processing in later stages.
The first stage of the back substitution launches $L$ blocks,
the next stage launches $L-1$ blocks, followed by $L-2$ blocks
in the third stage, etc.  So there are as many stages as the value of~$L$,
each stage launching one fewer block as the previous stage.

\section{Newton's method}

Given a system~$f(\x) = \zero$, with $\x = (x_1, x_2, \ldots, x_n)$,
we denote the matrix of all partial derivatives of~$f$ as $J_f$.
Given an initial approximation~$\x_0$ for a solution of~$f(\x) = \zero$,
the application of one step in Newton's method happens in two stages:
\begin{enumerate}
\item Evaluate $J_f$ and $f$ at~$\x_0$:
      $A = J_f(\x_0)$ and $\bfb = -f(\x_0)$.
\item Solve the linear system $A \Delta \x = \bfb$
      and update $\x_0$ to $\x_1 :=  \x_0 + \Delta \x$.
\end{enumerate}
Stating the stages explicitly as above we emphasize the separation
between the two stages in solving general polynomial systems where
the shape and structure of the polynomials varies widely between
almost linear to sparse systems with high degree monomials,
see for example the benchmark collection of PHCpack~\cite{Ver99}.

\section{Computational Results}

In this section we describe results with our preliminary implementations.
We selected two benchmark problems.  In the first, the cost of evaluation
and differentiation grows linearly in the dimension and the complexity of
Newton's method depends on the linear system solving.
In the second problem, also the cost of evaluation and differentiation
grows cubic in the dimension.

\subsection{hardware and software}

Our main target platform is the NVIDIA Tesla K20C,
which has 2496 cores with a clock speed of 706~MHz,
hosted by a Red Hat Enterprise Linux workstation of Microway,
with Intel Xeon E5-2670 processors at 2.6~GHz.
Our code was developed with version 4.4.7 of gcc
and version 5.5 of the CUDA compiler.

% Our other computer is an HP Z800 workstation 
% with 3.47 GHz Intel Xeon X5690,
% running Red Hat Enterprise Linux, hosting the
% NVIDIA Tesla C2050 has 448 cores at a clock speed of 1147 Mhz.
% The same compilers where used on both computers.

% We participated to the NVIDIA GPU Test Drive program of Microway
% and received access to a computer 
% with two 10-core Xeon E5-2680v2 2.8GHz CPUs
% and one NVIDIA Tesla Atlas GPU, running CentOs Linux 6.
% The version of the gcc compiler is 4.4.6
% and we used version 5.5 of the CUDA compiler.

The C++ code for the Gram-Schmidt method to run on the host is
directly based on the pseudo code and served mainly to verify
the correctness of our GPU code.
We compiled the programs with the optimization flag {\tt -O2}.
The code is available at github in the directory src/GPU
of PHCpack.

\subsection{The Chandrasekhar H-Equation}

The system arises from the discretization of an integral equation.
The problem was treated with Newton's method in~\cite{Kel80}
and added to a collection of benchmark problems in~\cite{Mor90}.
In~\cite{Gon95}, the system was studied with methods in computer algebra.
We follow the formulation in~\cite{Gon95}:
\begin{equation}
  \begin{array}{l}
   f_i(H_1,H_2,\ldots,H_n) \\
   ~~~= 2 n H_i - c H_i
{\displaystyle \left( \sum_{j=0}^{n-1} \frac{i}{i+j} H_j \right)}
     - 2n = 0, \\
   i=1,2,\ldots,n,
  \end{array}
\end{equation}
where $c$ is some real nonzero constant, $0 < c \leq 1$.
As we can write the equations for any dimension~$n$,
observe that the cost of evaluating the polynomials remains
linear in~$n$.  Also the cost of evaluating the columns of 
the Jacobian matrix linear in~$n$ as only the diagonal elements
contain $n$ linear terms.  The off-diagonal elements of the
Jacobian matrix consists of at most one linear term.
As the evaluation and differentiation cost for this problem
is linear in~$n$, this implies that the cost of one iteration
of Newton's method is dominated by the cost for solving the
linear system, which is cubic in~$n$.

Although the total number of solutions grows as $2^n$,
there is always one real solution with all its components positive
and relatively close to~1.  Starting at $H_i = 1$ for all $i$
leads to a quadratically convergent Newton's method.
The value for the parameter~$c$ we used in our experiments is $33/64$.
As all coefficients in the system and the solution are real,
the complex arithmetic is superfluous.  Nevertheless, we expect
the speedups to be the same if we would use only real arithmetic.

Although in our methodology, not taking advantage of
the shape and structure of the polynomial system,
it does not seem possible to obtain correct results without
the use of double double arithmetic, it may very well be
that the Jacobian matrix at the interesting solution is
diagonally dominant and that iterative methods in double arithmetic
will do very well to solve this particular benchmark problem.

To run Newton's method on this system, the experimental setup
is displayed in Figure~\ref{figsetupnew}.
\begin{figure}[hbt]
\begin{center}
\begin{tabbing}
for\= ~a number of iterations : \\
   \> 1. \= The host evaluates and differentiates the system \\
   \>    \> at the current approximation. \\
   \>    \> This result of the evaluation and differentiation \\
   \>    \> is stored in an $n$-by-$(n+1)$ matrix~$[A~\bfb]$, \\
   \>    \> with $\bfb = -f(H_1,H_2, \ldots, H_n)$. \\
   \>    \> The first component of $\bfb$ is printed. \\
   \> 2. \> $A \Delta \x = \bfb$ is solved 
            in the least squares sense, \\
   \>    \> either entirely by the host; or \\
   \>    \> if \= accelerated, then \\
   \>    \>    \> 2.1 \= the matrix $[A~\bfb]$ is transferred \\
   \>    \>    \>     \> from the host to the device; \\
   \>    \>    \> 2.2 the device does a QR decomposition on $[A~\bfb]$ \\
   \>    \>    \>     \> and back substitution on the 
                      system~$R \Delta \x = \y$;  \\
   \>    \>    \> 2.3 the matrices $Q$, $R$, and the solution $\Delta \x$ \\
   \>    \>    \>     \> are transferred from the device to the host. \\
   \> 3. \>  The host performs the update $\x = \x + \Delta \x$ \\
   \>    \>  to compute the new approximation. \\
   \>    \>  The first component of $\Delta \x$ and $\x$ are printed.
\end{tabbing}
\caption{Experimental setup to accelerate Newton's method.}
\label{figsetupnew}
\end{center}
\end{figure}

\begin{figure}[hbt]
\begin{center}
\epsfig{figure=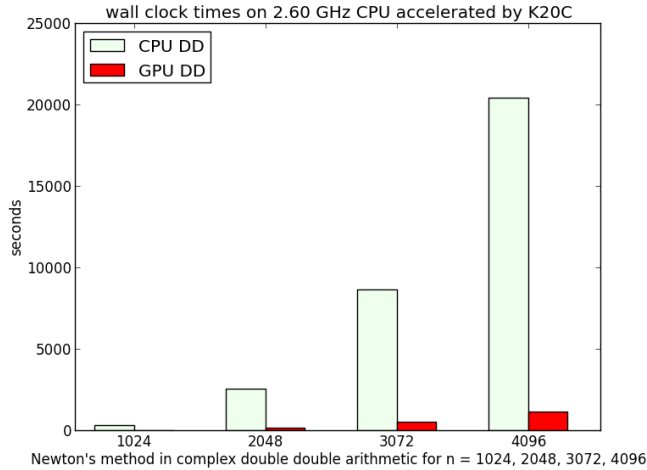,width=9cm}
\caption{Figure visualizing the data of Table~\ref{tabnewdd}.
Observe that the rightmost bar representing the accelerated run for $n=4096$
is less high than the bar for $n=2048$ without acceleration.
With acceleration we can double the dimension and still obtain
the results twice as fast.}
\label{fignewdd}
\end{center}
\end{figure}

Table~\ref{tabnewdd} with corresponding bar plot in Figure~\ref{fignewdd}
shows the running times obtained with the
command {\tt time} at the command prompt.
Comparing absolute real wall clock times: 
when we double the dimensions from 2048 to 4096,
the accelerated versions of the code run twice as fast,
20 minutes versus 42 minutes without acceleration.
As the cost of evaluation and differentiation grows only
linearly in the dimension, the cost of the linear solving
dominates and we see that as the dimension grows,
the difference in speedups between the two accelerated versions
fades out.

\begin{table}[hbt]
\begin{center}
\caption{Running six iterations of Newton's method in complex double double
  arithmetic on one core on the CPU and 
  accelerated by the K20C with block size equal to 128,
  once with the evaluation and differentiation done by the CPU (GPU1)
  and once with all computations on the GPU (GPU2).}
\label{tabnewdd}
\begin{tabular}{cl|rrr|c}
$n$ & mode  &      real~~  &    user ~~  &      sys~~ & speedup \\ \hline
1024 & CPU  &    5m22.360s &   5m21.680s &     0.139s & \\
     & GPU1 &      24.074s &     18.667s &     5.203s & 13.39 \\
     & GPU2 &      20.083s &     11.564s &     8.268s & 16.05 \\ \hline
2048 & CPU  &   42m41.597s &  42m37.236s &     0.302s & \\
     & GPU1 &    2m45.084s &   1m48.502s &    56.175s & 15.52 \\
     & GPU2 &    2m29.770s &   1m26.373s &  1m03.014s & 17.10 \\ \hline
3072 & CPU  &  144m13.978s & 144m00.880s &     0.216s & \\
     & GPU1 &    8m50.933s &   5m34.427s &  3m15.608s & 16.30 \\
     & GPU2 &    8m15.565s &   4m43.333s &  3m31.362s & 17.46 \\ \hline
4096 & CPU  &  340m00.724s & 339m27.019s &     0.929s & \\
     & GPU1 &   20m26.989s &  13m39.416s &  6m45.799s & 16.63 \\
     & GPU2 &   19m24.243s &  11m01.558s &  8m20.698s & 17.52
\end{tabular}
\end{center}
\end{table}

\begin{figure}[hbt]
\begin{center}
\epsfig{figure=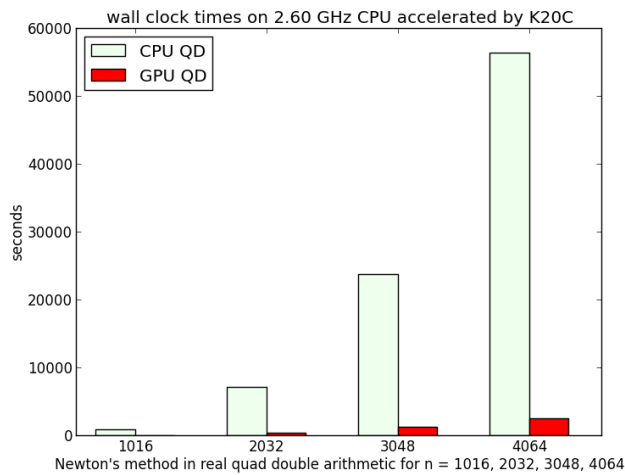,width=9cm}
\caption{Figure visualizing the data of Table~\ref{tabnewqd}.
Compared to complex double double arithmetic, we observe the timings
are of the same magnitude as the cost for complex double double is
similar to the cost of real quad double arithmetic.
For real quad doubles, the speedups are slightly better.}
\label{fignewqd}
\end{center}
\end{figure}

Table~\ref{tabnewqd} with corresponding bar plot in Figure~\ref{fignewqd}
shows the timings obtained from running
seven Newton iterations in real quad double arithmetic.
Because of shared memory limitations, the block size could not
be larger than~127 and our preliminary implementation requires
the dimension to be a multiple of the block size.

For quality up, we compare the {\tt 42m41.597s} in Table~\ref{tabnewdd}
for $n=2048$ with the {\tt 42m22.440s} in Table~\ref{tabnewqd}
for $n=4064$.  With the accelerated version we obtain twice as
accurate results for almost double the dimensions in about the same time.

\begin{table}[h!]
\begin{center}
\caption{Running seven iterations of Newton's method in real quad double
  arithmetic on one core on the CPU and 
  accelerated by the K20C (GPU) with block size equal to 127,
  with the evaluation and differentiation done by the CPU.}
\label{tabnewqd}
\begin{tabular}{cl|rrr|c}
$n$ & mode &      real~~  &    user ~~  &      sys~~ & speedup \\ \hline
1016 & CPU &   14m52.539s &  14m50.502s &     0.511s & \\
     & GPU &      59.570s &     48.923s &    10.377s & 14.98 \\ \hline
2032 & CPU &  118m20.789s & 118m10.189s &     0.134s & \\
     & GPU &    6m42.595s &   4m25.458s &  2m16.315s & 17.64 \\ \hline
3048 & CPU &  396m08.623s & 395m34.182s &     0.560s & \\
     & GPU &   21m21.047s &  13m49.481s &  7m29.744s & 18.55 \\ \hline
4064 & CPU &  939m16.703s & 937m50.275s &     0.941s & \\
     & GPU &   42m22.440s &   26m2.594s & 16m16.286s & 22.17 \\ 
\end{tabular}
\end{center}
\end{table}

\newpage
\subsection{the cyclic $n$-roots problem}

A system relevant to operator algebras is:

\begin{equation} \label{eqcyclicsys}
   \begin{cases}
   x_{0}+x_{1}+ \cdots +x_{n-1}=0 \\
   x_{0}x_{1}+x_{1}x_{2}+ \dots +x_{n-2}x_{n-1}+x_{n-1}x_{0}=0 \\
   i = 3, 4, \ldots, n-1: 
    \displaystyle\sum_{j=0}^{n-1} ~ \prod_{k=j}^{j+i-1}
    x_{k~{\rm mod}~n}=0 \\
   x_{0}x_{1}x_{2} \cdots x_{n-1} - 1 = 0. \\
\end{cases}
\end{equation}
The system is a benchmark problem in computer algebra:
\cite{AV13}, \cite{BF94}, \cite{DKK03}, \cite{Fau01}, \cite{Sab11}.

Except for the last equation, every polynomial has $n$ monomials,
so the total number of monomials grows as~$n^2 - n + 2$.
As each monomial is a product of at most $n$ variables,
the total cost to evaluate and differentiate the system is~$O(n^3)$.

\begin{figure}[hbt]
\begin{center}
\epsfig{figure=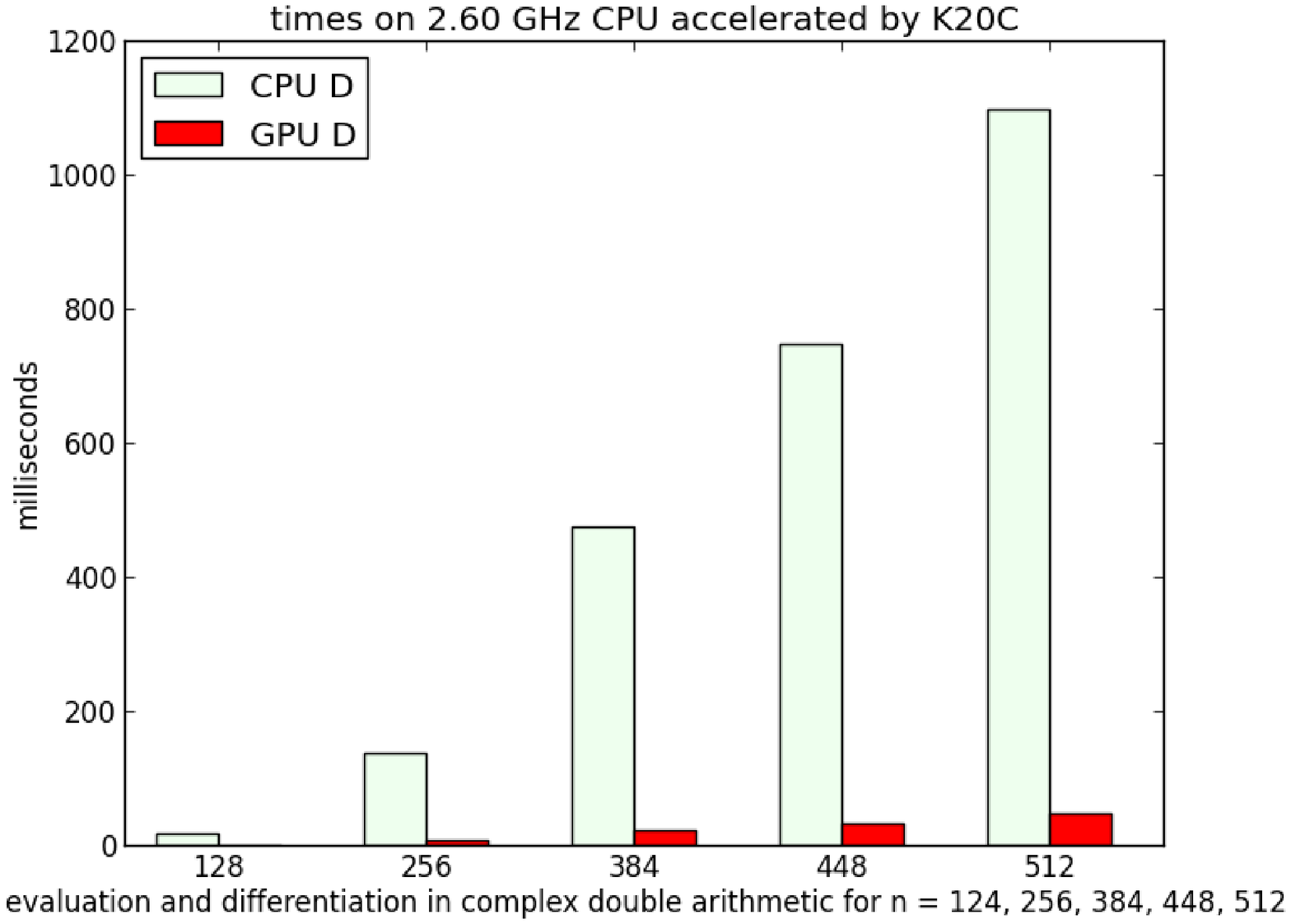,width=9cm}
\caption{Figure visualizing the data 
for complex double arithmetic (D) of Table~\ref{tabcyclicevaldiff}.
Observe that the rightmost bar representing the accelerated run 
for $n=512$ is about half the size of the bar
for $n=128$ without acceleration.
With acceleration we can double the dimension and still obtain
the results twice as fast.}
\label{figcyclicevaldiff}
\end{center}
\end{figure}

Table~\ref{tabcyclicevaldiff}
contains experimental results on the
evaluation of the polynomials in the cyclic $n$-roots system and
the evaluation of its Jacobian matrix.
For both the CPU and GPU we observe the $O(n^3)$ cost:
doubling the dimension increases the cost with a factor bounded by~8.
The speedups improve for larger problems and for increased precision,
see Figure~\ref{figcyclicevaldiff} for a plot of the results
in complex double arithmetic, using the algorithmic circuits
of Figures~\ref{figcircuit1} and~\ref{figcircuit2}.
For complex double double and quad double arithmetic,
the problem is no longer memory but is compute bound
and the computations in the DD and QD rows of 
Table~\ref{tabcyclicevaldiff} use global instead shared memory.

\begin{table}[h]
\begin{center}
\caption{Evaluation and differentiation of the cyclic $n$-roots problem
in complex double (D), complex double double (DD), 
and complex quad double (QD) arithmetic for increasing dimensions~$n$.
The times in milliseconds are obtained with the NVIDIA profiler.}
\label{tabcyclicevaldiff}
\begin{tabular}{rrr|r|r}
   & $n$ &     &    time~~~ & speedup \\ \hline
D  & 128 & CPU &    16.39ms &        \\
   &     & GPU &     1.13ms & 14.87~ \\
   & 256 & CPU &   136.26ms &        \\
   &     & GPU &     6.97ms & 19.55~ \\ 
   & 384 & CPU &   475.94ms &        \\
   &     & GPU &    21.59ms & 22.05~ \\
   & 448 & CPU &   747.44ms &        \\
   &     & GPU &    32.68ms & 22.87~ \\
   & 512 & CPU &  1097.20ms &        \\
   &     & GPU &    46.43ms & 23.63~ \\ \hline
DD & 128 & CPU &   144.27ms &        \\
   &     & GPU &     6.45ms & 22.36~ \\
   & 256 & CPU &  1169.07ms &        \\
   &     & GPU &    37.15ms & 31.47~ \\
   & 384 & CPU &  3981.07ms &        \\
   &     & GPU &   120.48ms & 33.04~ \\
   & 448 & CPU &  6323.17ms &        \\
   &     & GPU &   182.19ms & 34.52~ \\
   & 512 & CPU &  9411.55ms &        \\
   &     & GPU &   267.39ms & 35.20~ \\ \hline
QD & 128 & CPU &  1349.55ms &        \\
   &     & GPU &    29.45ms & 45.82~ \\
   & 256 & CPU & 10987.87ms &        \\
   &     & GPU &   152.82ms & 71.90~ \\
   & 384 & CPU & 37323.08ms &        \\
   &     & GPU &   513.78ms & 72.64~ \\
   & 448 & CPU & 59247.04ms &        \\
   &     & GPU &   809.15ms & 73.22~ \\
%      & \multicolumn{4}{|c}{Double}  \\ \hline
% CPU  & 16.39 & 136.27 & 475.94 & 747.44 & 1097.20 \\
% GPU & 1.13 &  6.97 &  21.59 &  32.68 &  46.43\\
%speedup & 14.87x  & 19.55x &  22.05x & 22.87x & 23.63x \\ \hline
%  & \multicolumn{4}{|c}{Double Double}  \\ \hline
%CPU & 144.27 & 1169.07 & 3981.07 & 6323.17 &  9411.55\\
%GPU & 6.45 &  37.15 &  120.48 &  183.19 & 267.39\\
%speedup &22.36x & 31.47x  & 33.04 & 34.52x & 35.20x \\ \hline
%&  \multicolumn{4}{|c}{Quad Double} \\ \hline
%CPU & 1349.55 & 10987.87 & 37323.08 & 59247.04\\
%GPU & 29.45 & 152.82  & 513.78 & 809.15\\
%speedup & 45.82x & 71.90x & 72.64x & 73.22x\\ \hline
\end{tabular}
\end{center}
\end{table}

We end this paper with the application of Newton's method
on the cyclic $n$-roots problem $f(\x) = \zero$ for $n=512$.
The setup is as follows.  We generate a random complex vector
$\z \in \cc^{512}$ and consider the system
$f(\x) - f(\z) t = \zero$, for $t < 1$.
For $t = 1$, we have that $\z$ is a solution
and for $t$ sufficiently close to~1, Newton's method will converge.
This setup corresponds to the start in running a Newton homotopy,
for $t$ going from one to zero.
In complex double double arithmetic, with seven iterations
Newton's method converges to the full precision.
The CPU time is 78,055.71 seconds
while the GPU accelerated time is 5,930.96 seconds,
reducing 21 minutes to about 1.6 minutes,
giving a speedup factor of about~13.

\section{Conclusions}

To accurately evaluate and differentiate polynomials in several variables
given in sparse distributed form we reorganized the arithmetic circuits
so all threads in block can contribute to the computation.
This computation is memory bound for double arithmetic and the
techniques to optimize a parallel reduction are beneficial also
for real double double arithmetic, but for complex double double
and quad double arithmetic the problem becomes compute bound.

We illustrated our CUDA implementation on two benchmark problems
in polynomial system solving.  For the first problem, the cost
of evaluation and differentiation grows linearly in the dimension
and then the cost of linear system solving dominates.
For systems with polynomials of high degree such as the
cyclic $n$-roots problem, the implementation to evaluate the
system and compute its Jacobian matrix achieved double digits speedups,
sufficiently large enough to compensate for one extra level of precision.
With GPU acceleration we obtain more accurate results faster,
for larger dimensions.

\section*{Acknowledgments} % use section* for acknowledgement

This material is based upon work supported 
by the National Science Foundation under Grant No.\ 1115777.
The Microway workstation with the NVIDIA Tesla K20C 
was purchased through a UIC LAS Science award.
% We gratefully acknowledge Microway for providing access to a
% Tesla-accelerated compute cluster.  In particular,
% the comparison between the K20C and the K40 were performed on 
% Microway's Tesla GPU accelerated compute cluster.
% http://www.microway.com/gpu-test-drive/

\bibliographystyle{plain}
% \bibliography{GPUnew}

\end{document}